\documentclass[MSNbibl,nameyear,dvips]{arxstspdf}
\usepackage{dcolumn}
\usepackage{graphicx}
\usepackage{flushend}
\usepackage{stfloats}

\volume{28}
\issue{2}
\pubyear{2013}
\firstpage{238}
\lastpage{256}
\doi{10.1214/13-STS414} 

\makeatletter

\newcolumntype{d}[1]{D{.}{.}{#1}}

\makeatother

\begin{document}
\begin{frontmatter}

\title{Handling Attrition in Longitudinal Studies: The Case for
Refreshment Samples}
\runtitle{Refreshment Samples and Attrition}

\begin{aug}
\author[a]{\fnms{Yiting} \snm{Deng}\ead[label=e1]{yiting.deng@duke.edu}},
\author[b]{\fnms{D. Sunshine} \snm{Hillygus}\ead[label=e2]{hillygus@duke.edu}},
\author[c]{\fnms{Jerome P.} \snm{Reiter}\corref{}\ead[label=e3]{jerry@stat.duke.edu}},
\author[d]{\fnms{Yajuan} \snm{Si}\ead[label=e4]{yajuan.si@columbia.edu}}
\and
\author[c]{\fnms{Siyu} \snm{Zheng}\ead[label=e5]{s.zheng@alumni.duke.edu}}
\runauthor{Y. Deng et al.}

\affiliation{Duke University}

\address[a]{Yiting Deng is Ph.D. Candidate, Fuqua School of Business, Duke University, Box 90120,
Durham, North Carolina
27708, USA \printead{e1}.\break}
\address[b]{D.~Sunshine Hillygus is Associate Professor, Department of Political
Science, Duke University, Box 90204, Durham, North Carolina
27708, USA \printead{e2}.}
\address[c]{Jerome P. Reiter is Professor and Siyu Zheng is B.S. Alumnus,
Department of Statistical Science, Duke University, Box 90251, Durham,
North Carolina 27708, USA \printead{e3}; \printead*{e5}.}
\address[d]{Yajuan Si is Postdoctoral Associate, Applied Statistical Center, Columbia
University, 1255 Amsterdam Avenue, New York,
New York 10027, USA \printead{e4}.}

\end{aug}

%
\begin{abstract}
Panel studies typically suffer from attrition, which
reduces sample size and can result in biased inferences.
It is impossible to know whether or not the attrition causes bias
from the observed panel data alone. Refreshment samples---new,
randomly sampled respondents given the questionnaire at the same time
as a subsequent wave of the panel---offer information that can be used
to diagnose and adjust for bias due to attrition. We review and bolster
the case for the use of refreshment samples in panel studies. We
include examples of both a fully Bayesian approach for analyzing the
concatenated panel and refreshment data, and a multiple imputation
approach for analyzing only the original panel. For the latter, we
document a positive bias in the usual multiple imputation variance
estimator. We present models appropriate for three waves and two
refreshment samples, including nonterminal attrition. We illustrate the
three-wave analysis using the 2007--2008 Associated Press--Yahoo! News
Election Poll.
\end{abstract}

%

\end{frontmatter}


\section{Introduction}\label{sec1}

Many of the the major ongoing government or government-funded surveys
have panel components including, for example, in the U.S., the American
National Election
Study (ANES), the General Social Survey (GSS), the Panel Survey on
Income Dynamics (PSID) and the Current Population Survey (CPS). Despite
the millions of dollars spent each year to
collect high quality data, analyses using panel data are inevitably
threatened by panel attrition (\cite{lynn2009methodology}), that is,
some respondents in the sample do not participate in later waves of
the study because they
cannot be located or refuse participation. For instance, the
multiple-decade PSID, first fielded in 1968,
lost nearly 50 percent of the initial sample members by 1989 due to
cumulative attrition and mortality. Even with a much shorter study
period, the 2008--2009 ANES Panel Study, which
conducted monthly interviews over the course of the 2008 election
cycle, lost 36 percent of respondents in less than a year.

At these rates, which are not atypical in large-scale
panel studies, attrition can have serious impacts on analyses that use
only respondents who completed all waves of the survey.
At best, attrition reduces effective sample size, thereby decreasing analysts'
abilities to discover longitudinal trends in behavior. At worst,
attrition results in an available sample\vadjust{\goodbreak} that is not representative of
the target population, thereby introducing potentially substantial
biases into
statistical inferences. It is not possible for analysts to determine
the degree to which attrition degrades complete-case analyses by using
only the collected data; external sources of information are needed.

One such source is refreshment samples. A refreshment sample includes
new, randomly sampled respondents who are given the
questionnaire at the same time as a second or subsequent wave of the
panel. Many of the large panel studies now routinely
include refreshment samples. For example, most of the
longer longitudinal studies of the National Center for Education
Statistics, including the Early\break Childhood Longitudinal Study and the
National Educational Longitudinal Study, freshened their samples at
some point in the study, either adding new panelists or as a separate
cross-section. The National Educational
Longitudinal Study, for instance, followed 21,500 eighth graders in two-year
intervals until 2000 and included refreshment samples in 1990 and
1992. It is worth noting that by the final wave of data
collection, just 50\% of the original sample remained in the panel.
Overlapping or rotating panel designs offer the equivalent of
refreshment samples.
In such designs, the sample is divided into
different cohorts with staggered start times such that one cohort of
panelists completes a follow-up interview at the same time another
cohort completes their baseline interview. So long as each cohort is
randomly selected and administered the same questionnaire, the
baseline interview of the new cohort functions as a refreshment
sample for the old cohort.
Examples of such rotating panel designs include the GSS and the Survey
of Income and Program Participation.


Refreshment samples provide information that can be used to assess the
effects of panel attrition and to correct for biases
via statistical modeling (\cite{hirano1998combining}). However, they
are infrequently used by analysts or data
collectors for these tasks.
In most cases, attrition is simply ignored,
with the analysis run only on those respondents who completed all
waves of the study (e.g., \cite{jelicic2009use}), perhaps with the
use of post-stratification weights
(\cite{vandecasteele2007attrition}).
This is done despite
widespread recognition among subject matter experts about the potential
problems of panel
attrition (e.g., \cite{ahern2005methodological}).

In this article, we review and bolster the case for the use of
refreshment samples in panel studies. We begin in Section~\ref{sec2}\vadjust{\goodbreak} by
briefly describing existing approaches for handling attrition that do
not involve refreshment samples. In Section~\ref{sec3} we present a
hypothetical two-wave panel to illustrate how refreshment samples can
be used to remove bias from nonignorable attrition.
In Section~\ref{sec4} we extend current models for refreshment samples, which are
described exclusively with two-wave settings in the literature, to
models for three waves and two refreshment samples. In doing so, we discuss
modeling nonterminal attrition in these settings, which arises when
respondents fail to respond to one wave but return to the study for a
subsequent one.
In Section~\ref{sec5} we illustrate the three-wave analysis using the
2007--2008 Associated Press--Yahoo! News Election Poll (APYN), which is a
panel study of the 2008 U.S.
Presidential election. Finally, in Section~\ref{sec6} we discuss some
limitations and open research issues in the use of refreshment samples.


\section{Panel Attrition in Longitudinal Studies}\label{sec2}



Fundamentally, panel attrition is a problem of nonresponse, so it is
useful to frame the various approaches to handling panel attrition
based on the assumed missing data mechanisms (\cite{rubin1976};
\cite{littlerubin}). Often
researchers ignore panel attrition entirely and use only the available
cases for analysis, for example, listwise deletion to create a balanced
subpanel (e.g., \cite{bartels1993messages};\break \cite{wawro2002estimating}).
Such approaches assume that the panel attrition is missing
completely at random\break (MCAR), that is, the missingness is
independent of observed and unobserved data. We speculate that this
is usually assumed for convenience, as often listwise deletion analyses
are not presented with empirical justification of MCAR
assumptions. To the extent that diagnostic analyses of MCAR
assumptions in panel attrition are conducted,
they tend to be reported and published separately from the substantive
research
(e.g., \cite{zabel1998analysis};
\cite{fitzgerald1998analysis}; \cite{bartels1999panel}; \cite
{clinton2001panel};
\cite{kruse2009panel}),
so that it is not clear if and how the diagnostics influence
statistical model specification.


%
%


Considerable research has documented that some individuals are more
likely to drop
out than others (e.g., \cite{behr2005extent}; \cite
{olsen2005problem}), making
listwise deletion a risky analysis strategy. Many \mbox{analysts} instead
assume that the data are missing at random (MAR), that is, missingness
depends on observed, but not unobserved, data.
One widely used MAR approach is to\vadjust{\goodbreak} adjust survey\break
weights for nonresponse, for example, by using
post-stratification
weights provided by the survey organization
(e.g., Henderson, Hillygus and Tompson\break (\citeyear{henderson2010sour})).
Re-weighting approaches assume that drop\-out occurs randomly within
weighting classes defined by observed variables that are associated
with dropout.

Although re-weighting can reduce bias introduced
by panel attrition, it is not a fail-safe solution.
There is wide variability in the way weights are
constructed and in the variables used. Nonresponse weights are often
created using demographic benchmarks, for example, from the CPS, but
demographic variables alone are unlikely to be adequate to correct for attrition
(\cite{vandecasteele2007attrition}). As is the case in other nonresponse
contexts, inflating weights can result in increased standard errors and
introduce instabilities due to particularly large or small weights
(\cite{lohr1999sampling}; \cite{gelman2007struggles}).

A related MAR approach uses predicted proba\-bilities
of nonresponse, obtained by modeling the \mbox{response} indicator as a
function of observed variables, as inverse probability weights to
enable inference by generalized estimating equations
(e.g., \cite{robinsrot95}; \cite{robinsrotzha}; Scharfstein, Rotnitzky and
  Robins\break (\citeyear{scharfrobrot});
\cite{chenyicook}). This potentially offers some robustness to model
misspecification, at~least asymptotically for MAR mechanisms, although
inferences can be
sensitive to large weights. One also can test whether or not parameters differ
significantly due to attrition for cases with complete data and cases
with incomplete data
(e.g., \cite{diggle89}; \cite{chenlittle}; \cite{qu02}; \cite{qu11}),
which can offer insight into the appropriateness of the assumed MAR
mechanism.\looseness=1

An alternative approach to re-weighting is single imputation, a method
often applied by statistical agencies in
general item nonresponse contexts (\cite{kalton1986treatment}). Single
imputation methods replace each missing value with a plausible guess,
so that the full panel can
be analyzed as if their data were complete. Although there are a wide range
of single imputation methods (hot deck, nearest neighbor, etc.)
that have been applied to missing data problems, the
method most specific to longitudinal studies is the
last-observation-carried-forward approach, in which an individual's
missing data are imputed
to equal his or her response in previous waves (e.g., \cite{packer1996double}).
Research has shown that this\vadjust{\goodbreak} approach can introduce substantial
biases in inferences (e.g., see \cite{hogandaniels}).

Given the well-known limitations of single imputation methods
(\cite{littlerubin}), multiple imputation (see Section~\ref{sec3})
also has been used to handle missing data from attrition
(e.g., \cite{pasek2009determinants}; \cite{honaker2010missing}). As
with the
majority of available methods used to correct for panel attrition,
standard approaches to multiple imputation assume an ignorable missing
data mechanism. Unfortunately, it is often expected that panel
attrition is not missing at random (NMAR), that is, the
missingness depends on unobserved data. In such cases, the only way
to obtain unbiased estimates of parameters is to model the
missingness. However, it is generally impossible to know the
appropriate model for
the missingness mechanism from the panel sample alone
(\cite{kristman2005methods}; \cite{basic2007assessing}; \cite{molenbeunk}).

Another approach, absent external data, is to handle the attrition directly
in the statistical models used for longitudinal data analysis
(\cite{verbmol}; \cite{diggleheag}; \cite{fitzmaurice};
\cite{hedeker}; \cite{hogandaniels}).
Here, unlike with other approaches, much research has focused on
methods for handling nonignorable
panel attrition. Methods include variants of
both selection models (e.g., \cite{hausman1979attrition};
\cite{siddiqui1996factors}; \cite{kenward}; \cite{scharfrobrot};
\cite{vella1999two}; \cite{das2004simple};
\cite{wooldridge2005simple}; \cite{semykina2010estimating}) and pattern
mixture models (e.g., \cite{littlepatmix};
\cite{kenwardmolenthijs}; \cite{roy}; \cite{linmccol}; Roy and\break Daniels (\citeyear{roydaniels})).
These model-based methods have to make untestable and typically
strong assumptions about the attrition process, again because there is
insufficient information in
the original sample alone to learn the missingness mechanism. It is
therefore prudent for analysts to examine how sensitive results
are to different sets of assumptions about attrition. We note that
\citet{rotrobschar98} and \citet{scharfrobrot} suggest related
sensitivity analyses for estimating equations with inverse probability
weighting.



\section{Leveraging Refreshment Samples}\label{sec3}



Refreshment samples are available in many panel studies, but the way
refreshment samples
are currently used with respect to panel attrition varies
widely. Initially, refreshment samples, as the name implies, were
conceived as a way to directly replace units who had dropped out (\cite
{ridder1992empirical}). The
general idea of using survey or field substitutes to correct for
nonresponse dates to some of the earliest survey methods work
(\cite{kish1959replacement}). Research has shown, however, that
respondent substitutes are more likely to
resemble respondents rather than nonrespondents, potentially
introducing bias without additional adjustments
(\cite{lin1995using}; \cite{vehovar1999field}; \cite{rubin2001using};
\cite{dorsett10}).
Also potentially problematic is when refreshment respondents are simply
added to the
analysis to boost the sample size, while the attrition
process of the original respondents is disregarded
(e.g., \cite{wissen1989dutch}; \cite{heeringa1997russia}; \cite
{thompson2006methods}).
In recent years, however, it is most common to see refreshment samples
used to diagnose panel attrition characteristics
in an attempt to justify an ignorable missingness assumption or as
the basis for discussion about potential bias in the results, without
using them for
statistical correction of the bias (e.g., \cite{Frick2006}; \cite
{kruse2009panel}).

Refreshment samples are substantially more powerful than suggested by
much of their previous use. Refreshment samples can be mined for
information about the attrition process, which in turn facilitates
adjustment of inferences for the missing data
(Hirano et~al., \citeyear{hirano1998combining,hirano2001combining}; \cite
{bartels1999panel}).
For example, the data can be used to construct inverse probability
weights for the cases in the panel (\cite{hirano1998combining}; \cite
{nevo2003using}), an approach we do not focus on here. They also offer
information for model-based
methods and multiple imputation (\cite{hirano1998combining}), which we
now describe and illustrate in detail.



\subsection{Model-Based Approaches}\label{sec3.1}

\begin{figure}

\includegraphics{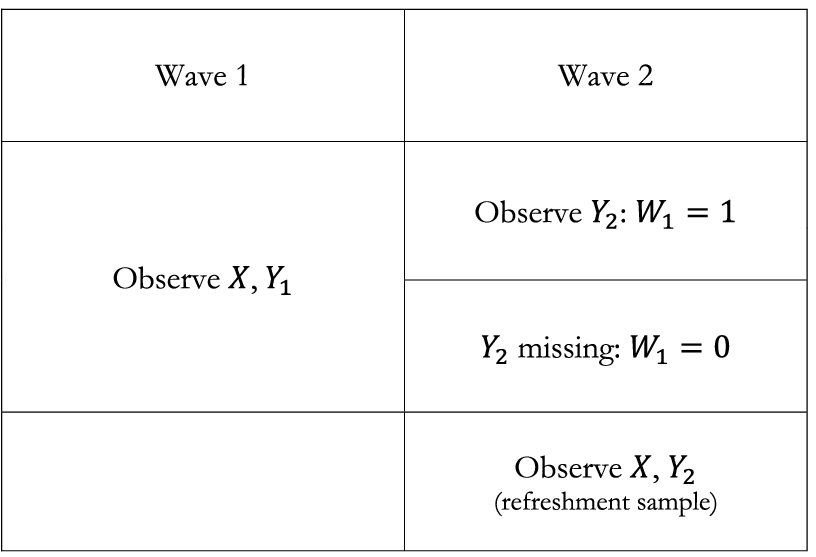}

\caption{Graphical representation of the two-wave
model. Here, $X$ represents variables
available on everyone.}\label{fig2-period}\vspace*{-3pt}
\end{figure}

Existing model-based methods for using refreshment samples
(\cite{hirano1998combining}; \cite{bhattacharya2008inference}) are
based on
selection models for the attrition process. To our knowledge, no one has
developed pattern mixture models in the context of refreshment
samples, thus, in what follows we only discuss selection models.
To illustrate these approaches, we use the simple example also
presented by
Hirano et~al. (\citeyear{hirano1998combining,hirano2001combining}),
which is
illustrated graphically in
Figure~\ref{fig2-period}. Consider a two-wave panel of $N_P$
subjects that includes a refreshment sample of $N_R$ new subjects
%
during the
second wave.
Let $Y_1$ and $Y_2$ be binary responses\vadjust{\goodbreak} potentially available in wave
1 and wave
2, respectively. For the original panel, suppose that we know $Y_1$
for all $N_P$ subjects and that we know $Y_2$ only for $N_{\mathit{CP}} < N_P$
subjects due to attrition. We also know $Y_2$ for the $N_R$ units in
the refreshment sample, but by design we do not know $Y_1$ for those
units. Finally, for all $i$, let $W_{1i} = 1$ if subject $i$ would
provide a value for $Y_2$ if they were included in wave 1, and let
$W_{1i}=0$ if subject $i$ would not provide a value for $Y_2$ if they
were included in wave 1. We note that $W_{1i}$ is observed for all $i$
in the original panel but is missing for all $i$ in the refreshment
sample, since they were not given the chance to respond in wave 1.

The concatenated data can be conceived as a partially observed,
three-way contingency table with eight cells. We can estimate the
joint probabilities in four of these cells from the observed data,
namely, $P(Y_1=y_1, Y_2=y_2,
W_1 =1)$ for $y_1, y_2 \in\{0,1\}$. We also have the following
three independent constraints involving the cells not directly observed:
\begin{eqnarray*}
&&
1 - \sum_{y_1, y_2} P(Y_1 = y_1,
Y_2 = y_2, W_1 = 1) \\[-1pt]
&&\quad= \sum
_{y_1,
y_2} P(Y_1 = y_1, Y_2 =
y_2, W_1 = 0),
\\[-1pt]
&&
P(Y_1 = y_1, W_1=0) \\[-1pt]
&&\quad= \sum
_{y_2} P(Y_1 = y_1, Y_2 =
y_2, W_1=0),
\\[-1pt]
&&
P(Y_2= y_2) - P(Y_2=y_2,
W_1 = 1) \\[-1pt]
&&\quad= \sum_{y_1} P(Y_1
= y_1, Y_2 = y_2, W_1=0).
\end{eqnarray*}
Here, all quantities on the left-hand side of the equations are
estimable from the observed data. The system of\vadjust{\goodbreak} equations offers seven
constraints for eight cells, so that we must add one constraint to
identify all the joint probabilities.

%
\begin{table*}
\tablewidth=400pt
\caption{Summary of simulation study for the
two-wave example. Results include the average of the posterior
means across the 500 simulations and the percentage of the 500
simulations in which the 95\% central posterior interval covers the
true parameter value. The implied Monte Carlo standard error of the
simulated coverage rates is
approximately $\sqrt{(0.95)(0.05)/500} = 1$\%}\label{tabHWandANNMAR}
\begin{tabular*}{\tablewidth}{@{\extracolsep{\fill}}ld{2.1}d{2.2}cd{2.2}d{2.0}d{2.2}c@{}}
\hline
&& \multicolumn{2}{c}{\textbf{HW}}
& \multicolumn{2}{c}{\textbf{MAR}} & \multicolumn{2}{c@{}}{\textbf{AN}} \\
[-4pt]
&& \multicolumn{2}{l}{\hspace*{-1.5pt}\hrulefill}
& \multicolumn{2}{l}{\hspace*{-1.5pt}\hrulefill} & \multicolumn{2}{l@{}}{\hspace*{-1.5pt}\hrulefill} \\
\textbf{Parameter} & \multicolumn{1}{c}{\textbf{True value}} & \multicolumn{1}{c}{\textbf{Mean}}
& \multicolumn{1}{c}{\textbf{95\% Cov.}} & \multicolumn{1}{c}{\textbf{Mean}}
& \multicolumn{1}{c}{\textbf{95\% Cov.}} & \multicolumn{1}{c}{\textbf{Mean}}
& \multicolumn{1}{c@{}}{\textbf{95\% Cov.}} \\
\hline
$\beta_{0}$ &0.3 &0.29 & 96 &0.27 & 87 &0.30 & 97\\
$\beta_{X}$ & -0.4 & -0.39 & 95 & -0.39 & 95 & -0.40 & 96\\
$\gamma_{0}$ &0.3 &0.44 & 30 &0.54 & 0&0.30 & 98\\
$\gamma_{X}$ & -0.3 & -0.35 & 94 & -0.39 & 70 & -0.30 & 99\\
$\gamma_{Y_{1}}$ &0.7 &0.69 & 91 &0.84 & 40 &0.70 & 95\\
$\alpha_{0}$ & -0.4 & -0.46 & 84&0.25 & 0 & -0.40 & 97\\
$\alpha_{X}$ & 1 &0.96 & 93 &0.84 & 13 & 1.00 & 98\\
$\alpha_{Y_{1}}$ & -0.7 & \multicolumn{1}{c}{---} & --- & -0.45 & 0 & -0.70 & 98\\
$\alpha_{Y_{2}}$ & 1.3 &0.75 & 0 & \multicolumn{1}{c}{---} & \multicolumn{1}{c}{---} & 1.31& 93\\
\hline
\end{tabular*}
\end{table*}

Hirano et~al. (\citeyear{hirano1998combining,hirano2001combining}) suggest
characterizing the joint
distribution of $(Y_1, Y_2, W_1)$ via a chain of conditional models, and
incorporating the additional constraint within the modeling
framework. In this context, they suggested letting
%
\begin{eqnarray}
\label{equ1}
Y_{1i} &\sim& \operatorname{Ber}(\pi_{1i}),\nonumber\\[-9pt]\\[-9pt]
\operatorname{logit}(\pi_{1i}) &=&
\beta_0,\nonumber
\\[-2pt]
\label{equ2}
Y_{2i}|Y_{1i} &\sim& \operatorname{Ber}(\pi_{2i}),\nonumber\\[-9pt]\\[-9pt]
\operatorname{logit}(\pi_{2i}) &=& \gamma _0+\gamma_1 Y_{1i},\nonumber
\\[-2pt]
\label{equ3}
W_{1i}|Y_{2i},Y_{1i} &\sim& \operatorname{Ber}
(\pi_{W_{1i}}),\nonumber\\[-8pt]\\[-8pt]
\operatorname{logit}(\pi _{W_{1i}})&=& \alpha_0+
\alpha_{Y_1}Y_{1i}+\alpha_{Y_2}Y_{2i}\nonumber
\end{eqnarray}
for all $i$ in the original panel and refreshment sample, plus
requiring that all eight probabilities sum to one. \citet
{hirano1998combining} call this an additive nonignorable
(AN) model. The AN model enforces the additional constraint by
disallowing the interaction between $(Y_1, Y_2)$ in (\ref{equ3}).
\citet{hirano1998combining} prove that the AN model is
likelihood-identified for general distributions. Fitting AN models is
straightforward using Bayesian MCMC; see \citet{hirano1998combining}
and \citet{dengMS}
for exemplary Metropolis-within-Gibbs algorithms. Parameters
also can be estimated via equations of moments (\cite
{bhattacharya2008inference}).

Special cases of the AN model are informative. By setting $(\alpha_{Y_2}=0,
\alpha_{Y_1} \ne0)$, we specify a model for a MAR missing data mechanism.
Setting $\alpha_{Y_2} \ne0$ implies a NMAR missing data
mechanism. In fact, setting $(\alpha_{Y_1} = 0, \alpha_{Y_2} \ne0)$
results in
the nonignorable model of \citet{hausman1979attrition}.\vadjust{\goodbreak} Hence,
the AN
model allows
the data to determine whether the missingness is MAR or NMAR, thereby
allowing the analyst to avoid making an untestable choice between the
two mechanisms. By not forcing $\alpha_{Y_1}=0$, the AN model permits
more complex
nonignorable selection mechanisms than the model of
\citet{hausman1979attrition}. The AN model does require separability
of $Y_1$ and $Y_2$ in the selection model; hence, if attrition depends
on the
interaction between $Y_1$ and $Y_2$, the AN model will not fully
correct for biases due to nonignorable attrition.


%


As empirical evidence of the potential of refreshment samples, we
simulate 500 data sets based on an extension of the model in
(\ref{equ1})--(\ref{equ3}) in which we add a Bernoulli-generated
covariate $X$ to each model; that is, we add $\beta_{X}X_i$ to the
logit predictor in (\ref{equ1}), $\gamma_X X_i$ to the logit predictor
in (\ref{equ2}), and $\alpha_X X_i$ to the logit predictor
in~(\ref{equ3}). In each we use $N_P=10\mbox{,}000$ original panel
cases and $N_R = 5000$ refreshment sample cases. The parameter values,
which are displayed in Table~\ref{tabHWandANNMAR}, simulate a
nonignorable missing data mechanism. All values of $(X, Y_1, W_1)$ are
observed in the original panel, and all values of $(X, Y_2)$ are
observed in the refreshment sample. We estimate three models based on
the data: the \citet{hausman1979attrition} model (set
$\alpha_{Y_1}=0$ when fitting the models) which we denote with HW, a
MAR model (set $\alpha_{Y_2}=0$ when fitting the models) and an AN
model. In each data set, we estimate posterior means and 95\% central
posterior intervals for each parameter using a Metropolis-within-Gibbs
sampler, running 10,000 iterations (50\% burn-in). We note that
interactions involving $X$ also can be included and identified in the
models, but we do not use them here.\vadjust{\goodbreak}


For all models, the estimates of the intercept and coefficient in the
logistic regression of $Y_1$ on $X$
are reasonable,
primarily because $X$ is complete and $Y_1$ is only MCAR in the
refreshment sample. As expected, the MAR model results in biased point
estimates and poorly calibrated intervals for the coefficients of the
models for $Y_2$ and $W_1$. The HW model fares somewhat better, but it
still leads to severely biased point estimates and poorly calibrated
intervals for $\gamma_0$ and $\alpha_{Y_2}$. In contrast, the AN
model results in approximately unbiased point estimates with
reasonably well-calibrated intervals.

We also ran simulation studies in which the data generation mechanisms
satisfied the HW and MAR models. When $(\alpha_{Y_1}=0, \alpha_{Y_2}
\neq0)$, the HW model performs well and the MAR model performs
terribly, as expected. When $(\alpha_{Y_1} \neq0, \alpha_{Y_2} = 0)$,
the MAR
model performs well and the HW model performs terribly, also as
expected. The AN model performs well in both scenarios, resulting in
approximately unbiased point estimates with reasonably well-cali\-brated
intervals.

To illustrate the role of the separability assumption, we repeat the
simulation study after including a nonzero interaction between $Y_1$
and $Y_2$ in the model for $W_1$. Specifically, we generate data
according to a response model,
%
\begin{eqnarray}\label{equ4}
\operatorname{logit}(\pi_{W_{1i}})&=&\alpha_0+\alpha_{Y_1}Y_{1i}+
\alpha_{Y_2}Y_{2i}\nonumber\\[-8pt]\\[-8pt]
&&{}+ \alpha_{Y_1Y_2} Y_{1i}Y_{2i},\nonumber
\end{eqnarray}
setting $\alpha_{Y_1Y_2} = 1$. However, we continue to use the AN model
by forcing $\alpha_{Y_1Y_2} = 0$ when estimating parameters. Table
\ref
{tab2waveint}
summarizes the results of 100 simulation runs, showing substantial
biases in all parameters except $(\beta_{0}, \beta_{X}, \gamma_X,
\alpha_X)$.
The estimates of $(\beta_0, \beta_X)$ are unaffected by using the wrong
value for $\alpha_{Y_1 Y_2}$,
since all the information about the relationship between $X$ and $Y_1$
is in the first
wave of the panel. The estimates of $\gamma_X$ and $\alpha_X$
are similarly unaffected because $\alpha_{Y_1 Y_2}$ involves only $Y_1$
(and not~$X$), which is controlled for in the regressions.
Table~\ref{tab2waveint} also displays the results when using
(\ref{equ1}), (\ref{equ2}) and (\ref{equ4}) with $\alpha_{Y_1Y_2}=1$; that is,
we set $\alpha_{Y_1Y_2}$ at its true value in the MCMC estimation and
estimate all other parameters. After accounting for separability, we
are able to recover all true parameter values.

\begin{table}
\caption{Summary of simulation study for the
two-wave example without
separability. The true selection model includes a nonzero
interaction between $Y_1$ and $Y_2$ (coefficient $\alpha_{Y_1Y_2}=1$).
We fit the AN model plus the AN model adding the interaction
term set at its true value. Results include the
averages of the posterior means and posterior standard errors across 100
simulations}\label{tab2waveint}
\begin{tabular*}{\tablewidth}{@{\extracolsep{\fill}}ld{2.1}d{2.2}cd{2.2}d{1.2}@{}}
\hline
&&\multicolumn{2}{c}{\textbf{AN}} &
\multicolumn{2}{c@{}}{$\bolds{\mathrm{AN} + \alpha_{Y_1Y_2}}$} \\
[-4pt]
&&\multicolumn{2}{l}{\hspace*{-1.5pt}\hrulefill} &
\multicolumn{2}{l@{}}{\hspace*{-1.5pt}\hrulefill}
\\
\textbf{Parameter} & \multicolumn{1}{c}{\textbf{True value}}
& \multicolumn{1}{c}{\textbf{Mean}} & \multicolumn{1}{c}{\textbf{S.E.}}
& \multicolumn{1}{c}{\textbf{Mean}} & \multicolumn{1}{c@{}}{\textbf{S.E.}} \\
\hline
$\beta_{0}$ &0.3 &0.30 &0.03 &0.30 &0.03\\
$\beta_{X}$ & -0.4 & -0.41 &0.04 & -0.41 &0.04\\
$\gamma_{0}$ &0.3 &0.14 &0.06&0.30 &0.06\\
$\gamma_{X}$ & -0.3 & -0.27 &0.06 & -0.30 &0.05\\
$\gamma_{Y_{1}}$ &0.7 &0.99 &0.07 &0.70 &0.06\\
$\alpha_{0}$ & -0.4 & -0.55 &0.08 & -0.41 &0.09\\
$\alpha_{X}$ & 1 &0.99 &0.08 & 1.01 &0.08\\
$\alpha_{Y_{1}}$ & -0.7 & -0.35 &0.05 & -0.70 &0.07\\
$\alpha_{Y_{2}}$ & 1.3 & 1.89 &0.13 & 1.31 &0.13\\
$\alpha_{Y_1Y_2}$ & 1 & \multicolumn{1}{c}{---} & \multicolumn{1}{c}{---} & 1 & 0 \\
\hline
\end{tabular*}        \vspace*{6pt}
\end{table}

Of course, in practice analysts do not know the
true value of $\alpha_{Y_1Y_2}$. Analysts who wrongly set\break $\alpha
_{Y_1Y_2}=0$, or any other incorrect value,
can expect bias patterns like those in Table~\ref{tab2waveint}, with
magnitudes determined by
how dissimilar the fixed $\alpha_{Y_1Y_2}$ is from the true value.
However, the successful recovery of true parameter values when setting
$\alpha_{Y_1Y_2}$ at its correct value
suggests an approach for analyzing the
sensitivity of inferences to the separability assumption. Analysts can
posit a set of plausible values for $\alpha_{Y_1Y_2}$, estimate the
models after fixing $\alpha_{Y_1Y_2}$ at each value and evaluate the
inferences that result. Alternatively, analysts might search for values of
$\alpha_{Y_1Y_2}$ that meaningfully alter substantive conclusions of
interest and judge whether or not such $\alpha_{Y_1Y_2}$ seem
realistic. Key to this sensitivity analysis is interpretation of
$\alpha_{Y_1Y_2}$. In the context of the model above,
$\alpha_{Y_1Y_2}$ has a natural interpretation in terms of odds
ratios; for example, in our simulation, setting $\alpha_{Y_1Y_2}=1$
implies that cases with $(Y_1 = 1, Y_2 =1)$ have $\exp(2.3) \approx10$
times higher odds of responding at wave~2 than cases with $(Y_1 = 1,
Y_2 = 0)$.
In a sensitivity analysis, when this is too high to seem realistic, we
might consider
models with values like $\alpha_{Y_1Y_2} =0.2$. Estimates from the
AN model can serve as starting points to facilitate interpretations.

Although we presented models only for binary data,
\citet{hirano1998combining} prove that similar models can be
constructed for other data types, for example, they present an analysis
with a
multivariate normal distribution for $(Y_1, Y_2)$. Generally speaking, one
proceeds by specifying a joint model for the outcome (unconditional on
$W_1$), followed by
a selection model for $W_1$ that maintains separation of $Y_1$ and $Y_2$.

\subsection{Multiple Imputation Approaches}\label{sec3.2}
\label{secMI}


One also can treat estimation with refreshment samples as a multiple
imputation exercise, in which one creates a modest number of completed
data sets to be analyzed with complete-data methods. In multiple
imputation, the basic idea is to simulate values for the missing data
repeatedly by sampling
from predictive distributions of the missing values. This creates $m>1$
completed data sets that can be analyzed or, as relevant for many statistical
agencies, disseminated to the public.
When the imputation models meet certain conditions (\cite{rubin1987},
Chapter 4), analysts of the $m$ completed data sets can obtain valid
inferences using complete-data statistical methods and software.
Specifically, the analyst computes point and variance estimates of
interest with each data set and combines these estimates using simple formulas
developed by \citet{rubin1987}. These formulas serve to
propagate the uncertainty introduced by missing values through the
analyst's inferences. Multiple imputation can be used for both MAR and
NMAR missing data,
although standard software routines primarily support MAR imputation schemes.
Typical approaches to multiple imputation presume either a joint model
for all the data, such as a multivariate normal or log-linear model
(\cite{schafer}), or use
approaches based on chained equations (Van~Buuren and Oudshoorn\break (\citeyear{oos}); \cite{raghu2001}).
See Rubin\break (\citeyear{rubin1996}), \citet{barnardmeng1999} and
\citet{reiterraghu07} for reviews of multiple imputation.

Analysts can utilize the refreshment samples when implementing multiple
imputation, thereby realizing similar benefits as illustrated in
Section~\ref{sec3.1}. First, the analyst fits the Bayesian models in
(\ref{equ1})--(\ref{equ3}) by running an MCMC algorithm for, say, $H$
iterations. This algorithm cycles between (i) taking draws of the
missing values, that is, $Y_2$ in the panel and $(Y_1,\break W_1)$ in the
refreshment sample, given parameter values and (ii) taking draws of the
parameters given completed data. After convergence of the chain, the
analyst collects $m$ of these completed data sets for use in multiple
imputation. These data sets should be spaced sufficiently so as to be
approximately independent, for example, by thinning the $H$ draws so
that the autocorrelations among parameters are close to zero. For
analysts reluctant to run MCMC algorithms, we suggest multiple
imputation via\break chained equations
with $(Y_1, Y_2, W_1)$ each taking turns as the dependent variable. The
conditional models
should disallow interactions (other than those involving $X$) to
respect separability.
This suggestion is based on our experience with limited simulation
studies, and we encourage further research into its
general validity. For the remainder of this article, we utilize the
fully Bayesian MCMC approach to
implement multiple imputation.

Of course, analysts could disregard the refreshment samples entirely
when implementing multiple imputation. For example, analysts can
estimate a MAR multiple imputation model by forcing $\alpha_{Y_2} = 0$
in (\ref{equ3}) and using the original panel only.
However, this model is exactly equivalent to the MAR model used in
Table~\ref{tabHWandANNMAR}
(although those results use both the panel and the refreshment sample
when estimating the model);
hence, disregarding the refreshment samples
can engender the types of biases and poor coverage rates observed
in Table~\ref{tabHWandANNMAR}. On the other hand, using the
refreshment samples allows the data to decide if MAR is appropriate or
not in the
manner described in Section~\ref{sec3.1}.

In the context of refreshment samples and the example in Section~\ref{sec3.1},
the analyst has two options for implementing multiple imputation. The
first, which we call the ``P${}+{}$R'' option, is to generate completed data
sets that include all cases for the
panel and refreshment samples, for example, impute the missing $Y_2$ in the
original panel and the missing $(Y_1, W_1)$ in the refreshment sample,
thereby creating $m$ completed data sets each with $N_P+N_R$ cases. The
second, which we call the ``P-only'' option, is to generate completed
data sets that include only
individuals from the initial panel, so that $N_P$ individuals are
disseminated or used for analysis. The estimation routines may
require imputing $(Y_1, W_1)$ for the refreshment sample cases, but in
the end only the imputed $Y_2$ are added to the observed data from the
original panel for dissemination/analysis.


For the P${}+{}$R option, the multiply-imputed data sets are byproducts when
MCMC algorithms are used to estimate the models.
The P${}+{}$R
option offers no advantage for analysts who would use the Bayesian
model for inferences, since essentially it just reduces from $H$ draws
to $m$ draws for summarizing
posterior distributions.
However, it could be useful for survey-weighted analyses, particularly
when the concatenated
file has weights that have been revised to reflect (as best as possible)
its representativeness. The analyst can apply
the multiple imputation methods of \citet{rubin1987} to the
concatenated file.\vadjust{\goodbreak}

Compared to the P${}+{}$R option, the P-only option offers clearer potential
benefits. Some statistical agencies or data analysts may find it easier to
disseminate or base inferences on only the original panel after using the
refreshment sample for imputing the missing values due to attrition,
since combining the original and freshened samples complicates
interpretation of sampling weights and design-based inference.
For example, re-weighting the concatenated data can be tricky with
complex designs in the original and refreshment sample. Alternatively,
there may be times when a statistical agency or other data collector
may not want to share the refreshment data with outsiders, for example,
because doing so would raise concerns over data confidentiality.
Some analysts might be reluctant to rely on the level of imputation in
the P${}+{}$R approach---for the refreshment sample, all $Y_1$ must be imputed.
In contrast, the P-only approach only leans on the imputation models
for missing $Y_2$. Finally, some analysts simply may prefer the
interpretation of longitudinal analyses based on the original panel,
especially in cases of multiple-wave designs.

In the P-only approach, the multiple imputation has a peculiar aspect:
the refreshment sample records used to estimate the imputation models
are not used or
available for analyses. When records are used for imputation but not for
analysis, \citet{reitermime} showed that Rubin's (\citeyear{rubin1987}) variance
estimator tends to have positive bias. The bias, which can be quite
severe, results from a mismatch in the conditioning used
by the analyst and the imputer. The derivation of Rubin's (\citeyear{rubin1987}) variance
estimator presumes that the analyst conditions on all records used in the
imputation models, not just the available data.

We now illustrate that this phenomenon also arises in the two-wave
refreshment sample context. To do so, we briefly review
multiple imputation (\cite{rubin1987}). For $l = 1, \ldots, m$, let
$q^{(l)}$ and $u^{(l)}$ be, respectively, the
estimate of some population quantity $Q$ and the estimate of the
variance of $q^{(l)}$ in
completed data set $D^{(l)}$. Analysts use $\bar{q}_{m} =
\sum_{l=1}^{m} q^{(l)}/m$ to estimate $Q$ and use $T_{m} = (1 + 1/m)
b_m + \bar{u}_{m}$
to estimate $\operatorname{var}(\bar{q}_{m})$, where $b_m = \sum_{l=1}^{m}
(q^{(l)} - \bar{q}_{m})^{2} / (m-1)$ and $\bar{u}_{m} = \sum_{l=1}^{m}
u^{(l)}/m$. For large samples, inferences for $Q$ are
obtained from the $t$-distribution, $(\bar{q}_m - Q) \sim t_{\nu_m}(0,
T_m)$, where the degrees of freedom is ${\nu_m} =
(m - 1)  [1+ \bar{u}_{m}/ ((1+1/m)b_m ) ]^{2}$.
A better
degrees of freedom for small samples is presented by \citet
{barnardrubin1999}.
Tests of significance for multicomponent null hypotheses are derived
by \citet{limengraghurub1991}, \citet{raghuetal1991},
\citet{mengrubin1992} and \citet{reitersmallndf}.


Table~\ref{tab1stepMI-2wave} summarizes the
properties of the P-only multiple
imputation inferences for the AN model under the simulation design
used for Table~\ref{tabHWandANNMAR}. We set $m=100$, spacing out
samples of parameters from the MCMC so as to have approximately
independent draws.
Results are based on 500 draws of observed data sets, each with new
values of missing data.
%
\begin{table}
\caption{Bias in multiple imputation
variance estimator for P-only method. Results based on 500 simulations}\label{tab1stepMI-2wave}
\begin{tabular*}{\tablewidth}{@{\extracolsep{\fill}}ld{2.1}d{2.2}ccc@{}}
\hline
\textbf{Parameter} & \multicolumn{1}{c}{$\bolds{Q}$} & \multicolumn{1}{c}{\textbf{Avg.} $\bolds{\bar{q}_*}$} &
\multicolumn{1}{c}{\textbf{Var} $\bolds{\bar{q}_*}$} & \multicolumn{1}{c}{\textbf{Avg.} $\bolds{T_*}$}
& \multicolumn{1}{c@{}}{\textbf{95\% Cov.}}\\
\hline
$ \beta_{0}$ &0.3 &0.30 &0.0008 &0.0008 & 95.4\\
$ \beta_{X}$ & -0.4 & -0.40 &0.0016 &0.0016 & 95.8\\
$ \gamma_{0}$ &0.3 &0.30 &0.0018 &0.0034 & 99.2\\
$ \gamma_{X}$ & -0.3 & -0.30 &0.0022 &0.0031 & 98.4\\
$ \gamma_{Y_1}$ &0.7 &0.70 &0.0031 &0.0032 & 96.4\\
\hline
\end{tabular*}
\end{table}
As before, the multiple imputation results in approximately unbiased
point estimates of the coefficients in the models for $Y_1$ and for
$Y_2$.
For the coefficients in the
regression of $Y_2$, the averages of $T_m$ across the 500 replications
tend to be significantly larger than the actual variances, leading to
conservative confidence interval coverage rates.
Results for the coefficients of $Y_1$ are well-calibrated; of course,
$Y_1$ has
no missing data in the P-only approach.

We also investigated the two-stage multiple imputation approach of
\citet{reitermime}. However, this resulted in some
anti-conservative variance
estimates, so that it was not preferred to standard multiple imputation.


\subsection{Comparing Model-Based and Multiple Imputation Approaches}\label{sec3.3}

As in other missing data contexts, model-based and multiple imputation
approaches have differential advantages (\cite{schafer}). For any
given model, model-based inferences tend to be more efficient than
multiple imputation inferences based on
modest numbers of completed data sets. On the other hand, multiple
imputation can be more robust than fully model-based approaches to
poorly fitting models. Multiple imputation uses the posited model
only for completing missing values, whereas a fully model-based
approach relies on the model for the entire inference. For example,
in the P-only approach, a~poor\-ly-specified imputation model
affects inference only through the $(N_P - N_{\mathit{CP}})$ imputations for $Y_2$.
Speaking loosely to offer intuition, if the model for $Y_2$ is only 60\%
accurate (a poor model indeed) and $(N_P - N_{\mathit{CP}})$ represents 30\%
of $N_P$, inferences based on the multiple imputations will be only
12\% inaccurate. In contrast, the full model-based inference will be
40\% inaccurate. Computationally,
multiple imputation has some advantages over model-based
approaches, in that analysts can use ad hoc imputation methods like
chained equations
(\cite{oos}; \cite{raghu2001}) that do not require MCMC.

Both the model-based and multiple imputation approaches, by
definition, rely on models for the data. Models that fail to describe
the data could result in inaccurate inferences, even when the
separability assumption in the selection model is reasonable. Thus,
regardless of the approach, it is prudent to check the fit of the
models to the observed data. Unfortunately, the literature on
refreshment samples
does not offer guidance on or present examples of such diagnostics.

We suggest that analysts check models with predictive distributions
(\cite{meng94ppp}; \cite{gelmanmengstern}; \cite{he2010}; \cite
{burgreit10}). In
particular, the analyst can use the estimated model to generate new
values of $Y_2$ for the complete cases in the original panel and for
the cases in the refreshment sample. The analyst compares
the set of replicated $Y_2$ in each sample with the corresponding
original $Y_2$ on statistics of interest, such as summaries of
marginal distributions and coefficients in regressions of $Y_2$ on observed
covariates. When the statistics from the replicated data and observed
data are dissimilar, the diagnostics indicate
that the imputation model does not generate replicated data that look
like the
complete data, suggesting that it may not describe adequately the relationships
involving $Y_2$ or generate plausible
values for the missing $Y_2$. When the statistics are similar, the
diagnostics do not offer evidence of imputation model inadequacy
(with respect to those statistics). We recommend that analysts generate
multiple sets of replicated data, so as to ensure interpretations are
not overly specific to particular replications.

These predictive checks can be graphical in nature, for example, resembling
grouped residual plots for logistic regression models. Alternatively,
as summaries analysts can compute posterior predictive
probabilities. Formally, let $S$ be the statistic of interest, such as a
regression coefficient or marginal probability. Suppose the analyst has
created\vspace*{1pt} $T$ replicated data sets, $\{R^{(1)}, \ldots, R^{(T)}\}$, where
$T$ is somewhat large (say,
$T=500$). Let $S_{D}$ and $S_{R^{(l)}}$ be the values of $S$ computed with
an observed subsample~$D$, for example, the complete cases in the panel
or the
refreshment sample, and $R^{(l)}$, respectively, where $l=1,\ldots,T$.
For each $S$ we compute the two-sided posterior predictive
probability,
%
\begin{eqnarray}\label{equ5}
\mathrm{ppp} &=& (2/T)*\min \Biggl(\sum_{l=1}^{T}I(S_{D}-S_{R^{(l)}}
> 0),\nonumber\\[-8pt]\\[-8pt]
&&\hspace*{64.5pt} \sum_{l=1}^{T}I(S_{R^{(l)}}-S_{D}
> 0) \Biggr).\nonumber
\end{eqnarray}
We note that $\mathrm{ppp}$ is small when $S_{D}$ and $S_{R^{(l)}}$ consistently
deviate from each
other in one direction, which would indicate that the
model is systematically distorting the relationship
captured by $S$. For $S$ with small $\mathrm{ppp}$, it is prudent to examine
the distribution of $S_{R^{(l)}} - S_{D}$ to evaluate if the
difference is practically important. We consider probabilities in the
0.05 range (or lower) as suggestive of lack of model fit.

To obtain each $R^{(l)}$, analysts simply add a step to the
MCMC that replaces all observed values of $Y_2$ using the
parameter values at that iteration, conditional on observed values of
$(X, Y_1, W_1)$. This step is used only to facilitate
diagnostic checks; the estimation of parameters continues to be
based on the observed $Y_2$. When autocorrelations among parameters
are high, we recommend thinning the chain so that parameter draws are
approximately independent before creating the set of
$R^{(l)}$. Further, we advise saving the $T$ replicated data sets, so
that they can be used repeatedly with different~$S$. We illustrate this
process of model
checking in the analysis of the APYN data in Section~\ref{sec5}.

\begin{figure*}

\includegraphics{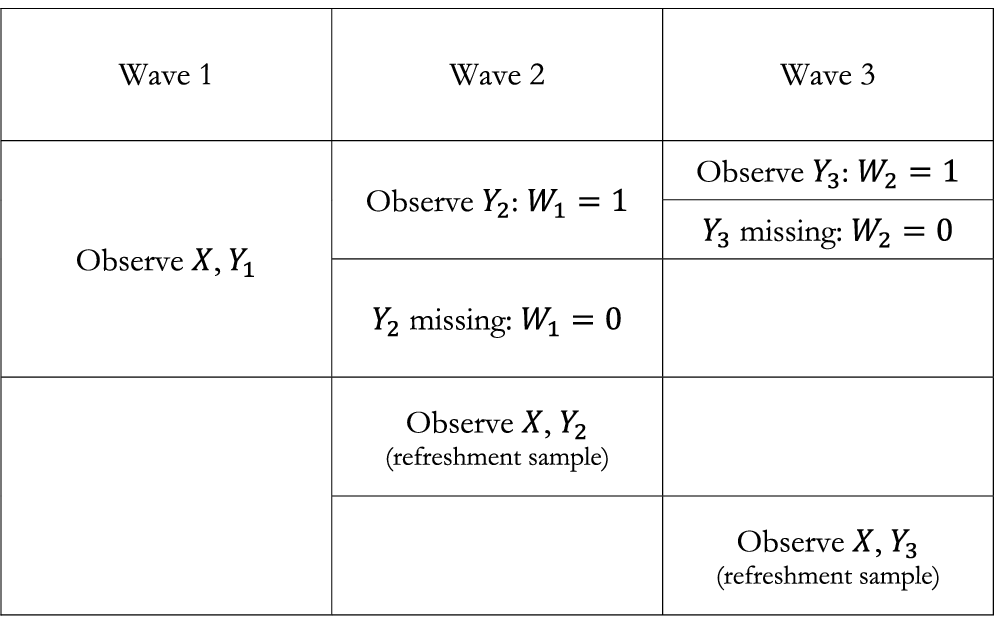}

\caption{Graphical representation of
the three-wave panel with monotone nonresponse and no follow-up
for subjects in refreshment samples. Here, $X$ represents variables
available on everyone and is displayed for generality; there is no $X$ in
the example in Section
\protect\ref{sec4}.}\label{fig3-period-monotone,nofollow}\vspace*{-6pt}
\end{figure*}


\section{Three-Wave Panels with Two Refreshments}\label{sec4}

To date, model-based and multiple imputation\break methods have been
developed and applied in the context of two-wave panel studies with
one refreshment sample. However, many panels exist for more than two
waves, presenting the opportunity for fielding multiple refreshment
samples under different designs. In this section we describe
models for three-wave panels with two refreshment
samples. These can be used as in Section~\ref{sec3.1} for model-based
inference or as in Section~\ref{sec3.2} to implement multiple imputation.
Model identification depends on (i) whether
or not individuals from the original panel who\vadjust{\goodbreak}
did not respond in the second wave, that is, have $W_{1i}=0$, are given the
opportunity to provide responses in the third wave, and (ii) whether
or not individuals from the first refreshment sample are followed in
the third wave.

To begin, we extend the example from Figure~\ref{fig2-period} to the case with
no panel returns and no refreshment follow-up, as illustrated in Figure
\ref{fig3-period-monotone,nofollow}.
Let $Y_3$ be binary responses potentially available in wave 3. For the
original panel, we know $Y_3$ only for $N_{\mathit{CP}2} < N_{\mathit{CP}}$
subjects due to third wave attrition. We also know $Y_3$ for the
$N_{R2}$ units in
the second refreshment sample. By design, we do not know $(Y_1, Y_3)$
for units in the first refreshment sample, nor do we know $(Y_1,
Y_2)$ for units in the second refreshment sample. For all $i$, let
$W_{2i} = 1$ if subject $i$ would
provide a value for $Y_3$ if they were included in the second wave of
data collection (even if they would not respond in that wave), and let
$W_{2i}=0$ if subject $i$ would not provide a value for $Y_3$ if they
were included in the second wave. In this design, $W_{2i}$ is missing
for all $i$
in the original panel with $W_{1i}=0$ and for all $i$ in both refreshment
samples.

There are 32 cells in the contingency table cross-tabulated from
$(Y_1, Y_2, Y_3, W_1, W_2)$. However, the observed data offer only sixteen
constraints, obtained from the eight joint probabilities
when $(W_1 = 1,\break W_2 = 1)$ and the following dependent equations\break (which
can be
alternatively specified). For all $(y_1, y_2,\break y_3, w_1, w_2)$, where
$y_3, w_1, w_2 \in\{0,1\}$, we have
\begin{eqnarray*}
&&1 = \sum_{y_1, y_2, y_3, w, w_2} P(Y_1 = y_1,
Y_2 = y_2,\\
&&\hspace*{86pt}Y_3 = y_3, W_1 = w_1, W_2 = w_2),
\\
&&
P(Y_1 = y_1, W_1 =0) \\
&&\quad= \sum
_{y_2, y_3, w_2} P(Y_1 = y_1, Y_2 =
y_2,\\
&&\hspace*{67.6pt} Y_3 = y_3, W_1 =0,
W_2 = w_2),
\\
&&
P(Y_2= y_2) - P(Y_2=y_2,
W_1 = 1) \\
&&\quad= \sum_{y_1, y_3, w_2} P(Y_1 =
y_1, Y_2 = y_2,\\
&&\hspace*{67.6pt}  Y_3 =
y_3, W_1=0, W_2=w_2),
\\
&&P(Y_1 = y_1, Y_2 = y_2,
W_1=1, W_2 = 0) \\
&&\quad= \sum_{y_3}
P(Y_1 = y_1, Y_2 = y_2,\\
&&\hspace*{50.4pt}
Y_3 = y_3, W_1 =1, W_2 = 0),
\\
&&P(Y_3=y_3) \\
&&\quad= \sum
_{y_1, y_2, w_1, w_2} P(Y_1 = y_1, Y_2 =
y_2,\\
&&\quad\hspace*{70.2pt} Y_3=y_3, W_1 =
w_1, W_2 = w_2).
\end{eqnarray*}
As before, all quantities on the left-hand side of the equations are
estimable from the observed data. The first three equations are
generalizations of those from the two-wave model. One can show
that the entire set of equations offers eight independent constraints,
so that we must add sixteen constraints to
identify all the probabilities in the table.

\begin{figure*}

\includegraphics{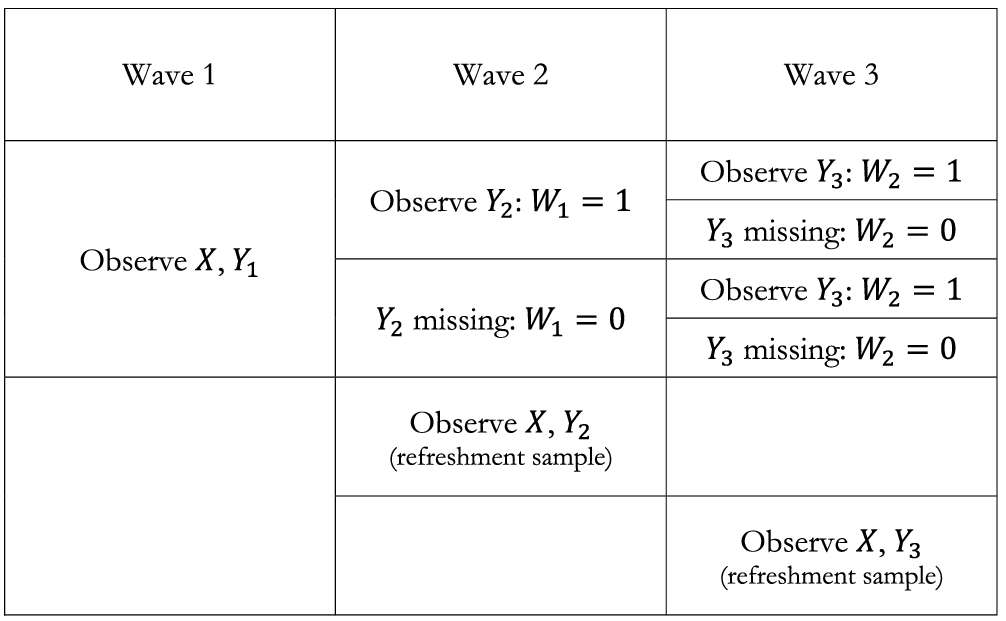}

\caption{Graphical representation of
the three-wave panel with return of wave 2 nonrespondents and no follow-up
for subjects in refreshment samples. Here, $X$ represents variables
available on everyone.}\label{fig3-period-nmonotone,nofollow}
\end{figure*}

Following the strategy for two-wave models, we characterize the joint
distribution of $(Y_1, Y_2, Y_3, W_1,\break W_2)$ via a chain
of conditional models. In particular, for all $i$ in the original
panel and refreshment samples, we supplement the models in
(\ref{equ1})--(\ref{equ3}) with
%
\begin{eqnarray}
\label{equ6}
Y_{3i} \mid Y_{1i}, Y_{2i}, W_{1i} &
\sim& \operatorname{Ber}(\pi_{3i}),
\nonumber\\
\operatorname{logit}(\pi_{3i}) &=& \beta_0+\beta_1
Y_{1i} \\
&&{}+ \beta_2 Y_{2i} + \beta_3
Y_{1i}Y_{2i},\nonumber
\\
\label{equ7}
\quad W_{2i} \mid Y_{1i}, Y_{2i}, W_{1i},
Y_{3i} &\sim& \operatorname{Ber}(\pi_{W2i}),
\nonumber\\
\operatorname{logit}(\pi_{W2i}) &=& \delta_0+\delta_1Y_{1i}+
\delta_2Y_{2i} \\
&&{}+ \delta_3 Y_{3i} +
\delta_4 Y_{1i}Y_{2i},
\nonumber
\end{eqnarray}
plus requiring that all 32 probabilities sum to one. We note that
the saturated model---which includes all eligible one-way,
two-way and three-way inter\-actions---contains 31
parameters plus the sum-to-one requirement, whereas the
just-identified model contains
15 parameters plus the sum-to-one requirement; thus, the needed 16 constraints
are obtained by fixing parameters in the saturated model to zero.

The sixteen removed terms from the saturated model include the
interaction $Y_1 Y_2$ from the model for $W_1$, all terms
involving $W_1$ from the model for $Y_3$ and all terms involving
$W_1$ or interactions with $Y_3$ from the model for $W_2$. We never
observe $W_1=0$ jointly with $Y_3$ or $W_2$, so that
the data cannot identify whether or not the distributions for $Y_3$ or
$W_2$ depend on $W_1$. We therefore require that $Y_3$ and
$W_2$ be conditionally independent of $W_1$. With this assumption,
the $N_{\mathit{CP}}$ cases with $W_1 = 1$ and the second refreshment sample can
identify the
interactions of $Y_1 Y_2$ in (\ref{equ6}) and (\ref{equ7}). Essentially, the $N_{\mathit{CP}}$
cases with fully observed $(Y_1, Y_2)$ and the second refreshment sample
considered in isolation are akin to a two-wave panel sample with $(Y_1,
Y_2)$ and their interaction as the variables from the ``first wave'' and
$Y_3$ as the variable from the ``second wave.'' As with the AN model, in
this pseudo-two-wave panel we can identify the
main effect of $Y_3$ in (\ref{equ7}) but not interactions involving $Y_3$.


In some multi-wave panel studies, respondents who complete the first
wave are invited to complete all subsequent waves, even if they
failed to complete a previous one. That is, individuals with observed
$W_{1i}=0$ can come back in future waves. For example, the 2008 ANES
increased incentives to attriters to
encourage them to return in later waves. This scenario is illustrated
in Figure~\ref{fig3-period-nmonotone,nofollow}. In such cases, the
additional information offers the potential to identify additional
parameters from the saturated model. In particular, one gains the
dependent equations,
\begin{eqnarray*}
&&
P(Y_1 = y_1, Y_3 = y_3,
W_1=0, W_2 = 1)\\
&&\quad= \sum_{y_2}
P(Y_1 = y_1, Y_2 = y_2,\\
&&\hspace*{51.5pt}
Y_3 = y_3, W_1 =0, W_2 = 1)
\end{eqnarray*}
for all $(y_1, y_3)$. When combined with other equations, we now
have 20 independent constraints. Thus, we can add four terms to the
models in (\ref{equ6}) and (\ref{equ7}) and maintain identification. These include two main
effects for $W_1$ and two interactions between $W_1$ and $Y_1$, all of
which are identified since we now observe some $W_2$ and $Y_3$ when
$W_1=0$. In contrast, the interaction term $Y_2 W_1$ is not identified, because
$Y_2$ is never observed with $Y_3$ except when $W_1=1$. Interaction
terms involving $Y_3$ also are not identified. This is intuitively\vadjust{\goodbreak}
seen by supposing that no values of $Y_2$ from the original panel were
missing, so that effectively the original panel plus the second
refreshment sample can be viewed as a two-wave setting in which the AN
assumption is required for $Y_3$.

Thus far we have assumed only cross-sectional refreshment samples,
however, refreshment sample respondents could be followed in
subsequent waves. Once again,
the additional information facilitates estimation of additional terms
in the models. For example, consider
extending Figure
\ref{fig3-period-nmonotone,nofollow} to include incomplete follow-up
in wave three for units from the first refreshment sample. \citet{dengMS}
shows that the observed data offer 22 independent
constraints, so that we can add six terms to (\ref{equ6}) and (\ref{equ7}). As before,
these include two main effects for $W_1$ and two interactions for
$Y_1W_1$. We also can add the two interactions for $Y_2 W_1$. The
refreshment sample follow-up offers observations with $Y_2$ and $(Y_3,
W_2)$ jointly observed, which combined with the other data enables
estimation of the one-way interactions. Alternatively, consider
extending Figure~\ref{fig3-period-monotone,nofollow} to include the
incomplete follow-up in wave three for units from the first refreshment sample.
Here, \citet{dengMS} shows that the observed data offer 20 independent
constraints and that one can add the two main effects for $W_1$ and two
interactions for $Y_2 W_1$ to (\ref{equ6}) and (\ref{equ7}).


As in the two-wave case (\cite{hirano1998combining}), we expect that
similar models can be constructed for other data types. We have done
simulation experiments (not reported here) that support this
expectation.

\section{Illustrative Application}\label{sec5}

To illustrate the use of refreshment samples in practice, we
use data from the 2007--2008 Associated Press--Yahoo! News Poll
(APYN). The APYN is a one year, eleven-wave survey with three
refreshment samples intended to measure attitudes about the 2008
presidential election and politics. The panel was sampled from the
probability-based Knowledge\-Panel(R) Internet panel, which recruits
panel members via a probability-based sampling method using known
published sampling frames that cover 99\% of the
U.S. population. Sampled noninternet households are provided a laptop
computer or MSN TV unit and free internet service.

The baseline (wave 1) of the APYN study was collected in November
2007, and the final wave took place after the November 2008 general
election. The baseline was fielded to a sample of 3548 adult
citizens, of\vadjust{\goodbreak} whom 2735 responded, for a 77\% cooperation
rate.
All baseline respondents were invited to participate in each follow-up
wave; hence, it is possible, for
example, to obtain a baseline respondent's values in wave $t+1$ even if they
did not participate in wave~$t$.
Cooperation rates in follow-up surveys varied from 69\%
to 87\%, with rates decreasing towards the end of the panel.
Refreshment samples were collected during follow-up waves in January,
September and October 2008. For illustration, we use only the data
collected in the baseline, January and October waves, including the
corresponding refreshment samples. We assume nonresponse to the
initial wave and to the refreshment samples is ignorable and analyze
only the available cases.
The resulting data set is akin to Figure
\ref{fig3-period-nmonotone,nofollow}.

\begin{table}[b]
\caption{Campaign interest. Percentage
choosing each response option across the panel waves (P1, P2, P3)
and refreshment samples (R2, R3). In P3,
83 nonrespondents from P2 returned to the survey. Five participants with
missing data in P1 were not used in the analysis}\label{tabcampaigninterest}
\begin{tabular*}{\tablewidth}{@{\extracolsep{4in minus 4in}}ld{2.1}d{2.1}d{2.2}d{2.1}d{2.1}@{}}
\hline
& \multicolumn{1}{c}{\textbf{P1}} & \multicolumn{1}{c}{\textbf{P2}}
& \multicolumn{1}{c}{\textbf{P3}} & \multicolumn{1}{c}{\textbf{R2}} &
\multicolumn{1}{c@{}}{\textbf{R3}}\\
\hline
``A lot'' & 29.8 & 40.3 & 65.0 & 42.0 & 72.2\\
``Some'' & 48.6 & 44.3 & 25.9 & 43.3 & 20.3\\
``Not much'' & 15.3 & 10.8 & 5.80 & 10.2 & 5.0\\
``None at all'' & 6.1 & 4.4 & 2.90 & 3.6 & 1.9\\
Available sample size & \multicolumn{1}{c}{2730} & \multicolumn{1}{c}{2316}
& \multicolumn{1}{c}{1715} & \multicolumn{1}{c}{691} & \multicolumn{1}{c@{}}{461}\\
\hline
\end{tabular*}
\end{table}

\begin{table*}
\tablewidth=410pt
\caption{Predictors used in all
conditional models, denoted as $X$. Percentage of respondents in
each category in initial panel (P1) and refreshment samples (R2,
R3)}\label{tabIndependent-Variables-Used}
\begin{tabular*}{\tablewidth}{@{\extracolsep{\fill}}llccc@{}}
\hline
\textbf{Variable} & \multicolumn{1}{c}{\textbf{Definition}} &
\multicolumn{1}{c}{\textbf{P1}} & \multicolumn{1}{c}{\textbf{R2}}
& \multicolumn{1}{c@{}}{\textbf{R3}}\\
\hline
AGE1 & $=$ 1 for age 30--44, $=$ 0 otherwise &0.28 &0.28 &0.21\\
AGE2 & $=$ 1 for age 45--59, $=$ 0 otherwise &0.32 &0.31 &0.34\\
AGE3 & $=$ 1 for age above 60, $=$ 0 otherwise &0.25 &0.28 &0.34\\
MALE & $=$ 1 for male, $=$ 0 for female &0.45 &0.47 &0.43\\
COLLEGE & $=$ 1 for having college degree, $=$ 0 otherwise &0.30 &0.33 &
0.31\\
BLACK & $=$ 1 for African American, $=$ 0 otherwise &0.08 &0.07 &0.07\\
INT & $=$ 1 for everyone (the intercept) & & & \\
\hline
\end{tabular*}
\end{table*}

The focus of our application is on campaign interest, one of the
strongest predictors of democratic attitudes and behaviors
(\cite{Prior2010}) and a key measure for defining likely voters in
pre-election polls (\cite{Traugott1984}). Campaign interest also has
been shown to be
correlated with panel attrition (\cite{bartels1999panel}; \cite{olson2011}).
For our analysis, we use an outcome variable
derived from answers to the survey question, ``How much thought, if
any, have
you given to candidates who may be running for president in 2008?''
Table~\ref{tabcampaigninterest} summarizes the distribution of
the answers in the three waves. Following convention
(e.g., \cite{pew2010}), we dichotomize answers into people most
interested in the campaign and all others. We let $Y_{ti}=1$ if subject
$i$ answers ``A lot'' at time $t$ and $Y_{ti}=0$ otherwise, where $t
\in\{1,2,3\}$ for the baseline, January and October waves,
respectively. We let $X_{i}$
denote the vector of predictors summarized in Table \ref
{tabIndependent-Variables-Used}.

We assume ignorable nonresponse in the initial wave and
refreshment samples for convenience, as our primary goal is
to illustrate the use and potential benefits of refreshment samples.
Unfortunately, we have little evidence in the data
to support or refute that assumption. We do not have access to
$X$ for the nonrespondents in the initial panel or refreshment
samples, thus, we cannot compare them to respondents' $X$ as a
(partial) test of an MCAR assumption. The
respondents' characteristics are reasonably
similar across the three samples---although the respondents in the
second refreshment sample (R3) tend to
be somewhat older than other samples---which offers some
comfort that, with respect to demographics, these three samples are not subject
to differential nonresponse bias.


As in Section~\ref{sec4}, we estimate a series of logistic regressions. Here,
we denote the $7 \times1$ vectors of coefficients in front of the
$X_i$ with $\theta$ and subscripts indicating the dependent
variable, for example, $\theta_{Y_1}$ represents the
coefficients of $X$ in the model for $Y_1$. Suppressing conditioning,
the series of models is
\begin{eqnarray*}
Y_{1i} &\sim& \operatorname{Bern} \biggl(\frac{\operatorname{exp} (\theta_{Y_1}X_{i}
)}{1+\operatorname{exp} (\theta_{Y_{1}} X_{i} )} \biggr),
\\
Y_{2i} &\sim& \operatorname{Bern} \biggl(\frac{\operatorname{exp} (\theta_{Y_2} X_{i}+\gamma
Y_{1i} )}{1+\operatorname{exp} (\theta_{Y_2} X_{i}+\gamma Y_{1i}
)} \biggr),
\\
W_{1i} &\sim& \operatorname{Bern} \biggl(\frac{\operatorname{exp} (\theta_{W_1}X_{i}+\alpha_{1}
Y_{1i}+\alpha_{2}
Y_{2i} )}{1+\operatorname{exp} (\theta_{W_1}X_{i}+\alpha_{1}Y_{1i}+\alpha
_{2}Y_{2i} )} \biggr),
\\
Y_{3i} &\sim& \operatorname{Bern} \bigl(\operatorname{exp} (\theta_{Y_3}
X_{i}+\beta_{1}Y_{1i}+\beta_{2}Y_{2i}\\
&&\hspace*{46pt}{}+\beta_{3}W_{1i}+\beta
_{4}Y_{1i}Y_{2i}+\beta_{5}Y_{1i}W_{1i} )\\
&&\hspace*{27pt}{}/\bigl(1+\operatorname{exp} (\theta_{Y_3}
X_{i}+\beta_{1}Y_{1i}\\
&&\hspace*{73pt}{}+\beta_{2}Y_{2i}+\beta_{3}W_{1i}\\
&&\hspace*{87pt}{}+\beta_{4}
Y_{1i}Y_{2i}+\beta_{5}Y_{1i}W_{1i} )\bigr) \bigr),
\\
W_{2i} &\sim& \operatorname{Bern}
\bigl(\operatorname{exp} (\theta_{W_2}X_{i}+\delta_{1}
Y_{1i}+\delta_{2}Y_{2i}+\delta_{3}Y_{3i}\\
&&\hspace*{43.7pt}{}+\delta_{4}W_{1i}+\delta
_{5}Y_{1i}Y_{2i}+\delta_{6}Y_{1i}W_{1i} )\\
&&\hspace*{26pt}{}/\bigl(1+\operatorname{exp} (\theta_{W_2}
X_{i}+\delta_{1}Y_{1i}+\delta_{2}Y_{2i}\\
&&\hspace*{71.3pt}{}+\delta_{3}Y_{3i}+\delta
_{4}W_{1i}\\
&&\hspace*{83.2pt}{}+\delta_{5}Y_{1i}Y_{2i}+\delta_{6}Y_{1i}W_{1i}
)\bigr) \bigr).
\end{eqnarray*}
We use noninformative prior distributions on all parameters. We
estimate posterior distributions of the parameters using a
Metropolis-within-Gibbs algorithm, running the chain for 200,000 iterations
and treating the first 50\% as burn-in. MCMC diagnostics
suggested that the chain converged. Running the MCMC for
200,000 iterations took approximately 3 hours on a standard desktop
computer (Intel Core 2 Duo CPU 3.00 GHz, 4 GB RAM). We developed the
code in Matlab without making significant efforts to optimize the
code. Of course, running times could be significantly faster with
higher-end machines and smarter coding in a language like C$++$.

\begin{table*}
\caption{Posterior means and 95\% central intervals
for coefficients in regressions.\break Column headers are the
dependent variable in the regressions}\label{tabY2est}
\begin{tabular*}{\tablewidth}{@{\extracolsep{\fill}}lccccc@{}}
\hline
\textbf{Variable} & \multicolumn{1}{c}{$\bolds{Y_1}$}
& \multicolumn{1}{c}{$\bolds{Y_2}$} & \multicolumn{1}{c}{$\bolds{Y_3}$}
& \multicolumn{1}{c}{$\bolds{W_1}$} & \multicolumn{1}{c@{}}{$\bolds{W_2}$}\\
\hline
INT & $-$1.60 & $-$1.77 &0.04 & 1.64 & $-$1.40\\
& $(-1.94, -1.28)$ & $(-2.21, -1.32)$ & $(-1.26, 1.69)$ & $(1.17, 2.27)$ &
$(-2.17, -0.34)$\\
AGE1 &0.25 &0.27 &0.03 & $-$0.08 &0.28\\
& $(-0.12,0.63)$ & $(-0.13,0.68)$ & $(-0.40,0.47)$ & $(-0.52,0.37)$ & $(-0.07,0.65)$\\
AGE2 &0.75 &0.62 &0.15 &0.24 &0.27\\
& $(0.40, 1.10)$ & $(0.24, 1.02)$ & $(-0.28,0.57)$ & $(-0.25,0.72)$ & $(-0.07,0.64)$\\
AGE3 & 1.26 &0.96 &0.88 &0.37 &0.41\\
& $(0.91, 1.63)$ & $(0.57, 1.37)$ & $(0.41, 1.34)$ & $(-0.14,0.87)$ & $(0.04,0.80)$\\
COLLEGE &0.11 &0.53 &0.57 &0.35 &0.58\\
& $(-0.08,0.31)$ & $(0.31,0.76)$ & $(0.26,0.86)$ & $(0.04,0.69)$ & $(0.34,0.84)$\\
MALE & $-$0.05 & $-$0.02 & $-$0.02 &0.13 &0.08\\
& $(-0.23,0.13)$ & $(-0.22,0.18)$ & $(-0.29,0.24)$ & $(-0.13,0.39)$ & $(-0.14,0.29)$\\
BLACK &0.75 & $-$0.02 &0.11 & $-$0.54 & $-$0.12\\
& $(0.50, 1.00)$ & $(-0.39,0.35)$ & $(-0.40,0.64)$ & $(-0.92,-0.14)$ & $(-0.47,0.26)$\\
$Y_{1}$ & --- & 2.49 & 1.94 &0.50&0.88\\
& --- & $(2.24, 2.73)$ & $(0.05, 3.79)$ & $(-0.28, 1.16)$ & $(0.20, 1.60)$\\
$Y_2$ & --- & --- & 2.03 & $-$0.58 &0.27\\
& --- & --- & $(1.61, 2.50)$ & $(-1.92,0.89)$ & $(-0.13,0.66)$\\
$W_1$ & --- & --- & $-$0.42 & --- & 2.47\\
& --- & --- & $(-1.65,0.69)$ & --- & $(2.07, 2.85)$\\
$Y_1Y_2$ & --- & --- & $-$0.37 & --- & $-$0.07\\
& --- & --- & $(-1.18,0.47)$ & --- & $(-0.62,0.48)$\\
$Y_1W_1$ & --- & --- & $-$0.52 & --- & $-$0.62\\
& --- & --- & $(-2.34, 1.30)$ & --- & $(-1.18, -0.03)$\\
$Y_3$ & --- & --- & --- & --- & $-$1.10\\
& --- & --- & --- & --- & $(-3.04, -0.12)$\\
\hline
\end{tabular*}
\end{table*}

The identification conditions include no interaction between
campaign interest in wave 1 and wave~2 when predicting attrition in
wave 2, and
no interaction between campaign interest in wave 3 (as well
as nonresponse in wave 2) and other variables when predicting attrition
in wave 3.
These conditions are impossible to check from the sampled data alone,
but we cannot think of any scientific basis to reject them outright.

Table~\ref{tabY2est} summarizes the posterior distributions of the
regression coefficients in each of the models.
Based on the model for $W_1$, attrition in the second wave is
reasonably described as
missing at random, since the coefficient of $Y_2$ in that model is not
significantly different from zero. The model for
$W_2$ suggests that attrition in wave 3 is not missing at random.
The coefficient for $Y_3$ indicates that participants who were
strongly interested in the election at wave 3 (holding all else
constant) were more likely to drop out. Thus, a panel attrition
correction is needed to avoid making biased inferences.

This result contradicts conventional wisdom that
politically-interested respondents are \textit{less} likely to attrite
(\cite{bartels1999panel}).
The discrepancy could result from differences in the survey design of
the APYN
study compared to previous studies with attrition. For example, the
APYN study
consisted of 10--15 minute online interviews, whereas the ANES panel
analyzed by \citet{bartels1999panel} and \citet{olson2011} consisted
of 90-minute, face-to-face interviews. The lengthy ANES interviews
have been linked to significant panel conditioning effects, in which
respondents change their attitudes and behavior as a result of
participation in the panel (\cite{bartels1999panel}). In contrast,
\citet{kruse2009panel} finds few panel conditioning effects in the
APYN study. More notably, there was a differential incentive
structure in the APYN study. In later waves of the study,
reluctant responders (those who took more than 7 days to respond
in earlier waves) received increased monetary incentives to
encourage their participation. Other panelists and the refreshment
sample respondents received\vadjust{\goodbreak} a standard incentive. Not surprisingly,
the less interested respondents were more likely to have
received the bonus incentives, potentially increasing their retention
rate to exceed that of the most interested respondents. This
possibility raises a broader question about the reasonableness of
assuming the initial nonresponse is ignorable, a point we return to in
Section~\ref{sec6}.

In terms of the campaign interest variables, the observed relationships with
$(Y_1, Y_2, Y_3)$ are consistent with previous research
(\cite{Prior2010}). Not surprisingly, the strongest predictor of
interest in later waves is interest in previous waves. Older and
college-educated participants are more likely to be interested in
the election. Like other analyses of the 2008 election
(\cite{lawless2009}), and in contrast to many previous election cycles,
we do not find a significant gender gap in campaign interest.

We next illustrate the P-only approach with multiple imputation.
We used the posterior draws of
parameters to create $m=500$ completed data sets of the original panel
only. We thinned the chains until autocorrelations of the parameters
were near zero to obtain the parameter sets. We then estimated
marginal probabilities of $(Y_2, Y_3)$ and a
logistic regression for $Y_3$ using maximum likelihood on only the
2730 original panel cases, obtaining inferences via Rubin's (\citeyear{rubin1987})
combining rules. For comparison, we estimated the same quantities using
only the 1632
complete cases, that is, people who completed all three waves.

The estimated marginal probabilities reflect the results
in Table~\ref{tabY2est}. There is little difference in $P(Y_2=1)$ in
the two analyses: the 95\% confidence interval is $(0.38,0.42)$ in the
complete cases and $(0.37,0.46)$ in the full panel after multiple
imputation. However,
there is a suggestion of attrition bias in $P(Y_3=1)$. The
95\% confidence interval is $(0.63,0.67)$ in the complete cases and
$(0.65,0.76)$ in the full panel after multiple imputation. The estimated $P(Y_3
=1 \mid W_2 = 0)=0.78$, suggesting that nonrespondents in the third
wave were substantially more interested in the campaign than
respondents.

Table~\ref{tabY3est-MI-2} displays the point estimates and 95\% confidence
intervals for the regression coefficients for both analyses.
The results from the two analyses are quite similar except for the
intercept, which is smaller after adjustment for attrition. The
relationship between a
college education and political interest is somewhat attenuated after
correcting for attrition,
although the confidence intervals in the two analyses overlap substantially.
Thus, despite an apparent attrition bias affecting the marginal
distribution of political interest in wave 3,
the coefficients for this particular complete-case analysis appear not
to be degraded by panel attrition.

\begin{table}
\caption{Maximum likelihood estimates and 95\% confidence
intervals based for coefficients of predictors of $Y_{3}$ using $m=500$
multiple imputations and only complete cases at final wave}\label{tabY3est-MI-2}
\begin{tabular*}{\tablewidth}{@{\extracolsep{4in minus 4in}}lll@{\hspace*{-1pt}}}
\hline
\textbf{Variable} & \multicolumn{1}{c}{\textbf{Multiple imputation}}
& \multicolumn{1}{c@{}}{\textbf{Complete cases}}\\
\hline
INT & $-$0.22 $(-0.80,0.37)$ & $-$0.64 $(-0.98, -0.31)$\\
AGE1 & $-$0.03 $(-0.40,0.34)$ & \hphantom{$-$}0.01 $(-0.36,0.37)$\\
AGE2 & \hphantom{$-$}0.08 $(-0.30,0.46)$ & \hphantom{$-$}0.12 $(-0.25,0.49)$ \\
AGE3 & \hphantom{$-$}0.74 $(0.31, 1.16)$ & \hphantom{$-$}0.76 $(0.36, 1.16)$\\
COLLEGE & \hphantom{$-$}0.56 $(0.27,0.86)$ & \hphantom{$-$}0.70 $(0.43,0.96)$\\
MALE & $-$0.09 $(-0.33,0.14)$ & $-$0.08 $(-0.32,0.16)$ \\
BLACK & \hphantom{$-$}0.07 $(-0.38,0.52)$ & \hphantom{$-$}0.05 $(-0.43,0.52)$\\
$Y_{1}$ & \hphantom{$-$}1.39 $(0.87, 1.91)$ & \hphantom{$-$}1.45 $(0.95, 1.94)$ \\
$Y_{2}$ & \hphantom{$-$}2.00 $(1.59, 2.40)$ & \hphantom{$-$}2.06 $(1.67, 2.45)$\\
$Y_{1}Y_{2}$ & $-$0.33 $(-1.08,0.42)$ & $-$0.36 $(-1.12,0.40)$ \\
\hline
\end{tabular*}
\end{table}

Finally, we conclude the analysis with a diagnostic check of the
three-wave model following the approach outlined in Section~\ref{sec3.3}.
To do so, we generate 500 independent replications of $(Y_2, Y_3)$
for each of the cells in Figure
\ref{fig3-period-nmonotone,nofollow} containing observed responses.
We then compare the estimated probabilities for $(Y_2, Y_3)$ in the
replicated data to the
corresponding probabilities in the observed data, computing the value
of $\mathrm{ppp}$ for each cell. We also estimate
the regression from Table~\ref{tabY3est-MI-2} with the replicated data
using only the complete cases in the panel, and compare
coefficients from the replicated data to those estimated with the complete
cases in the panel. As shown in Table~\ref{ppp}, the imputation models
generate data
that are highly compatible with the observed data in the panel and the
refreshment samples on both the
conditional probabilities and regression coefficients. Thus, from
these diagnostic checks we do not have evidence of lack of model fit.

\begin{table}
\caption{Posterior predictive probabilities ($\mathrm{ppp}$) based on 500
replicated data sets and various observed-data quantities. Results
include probabilities
for cells with observed data and coefficients in regression of $Y_3$
on several predictors estimated with complete cases in the panel}\label{ppp}
\begin{tabular*}{\tablewidth}{@{\extracolsep{\fill}}lc@{}}
\hline
\textbf{Quantity}& \multicolumn{1}{c@{}}{\textbf{Value of} $\bolds{\mathrm{ppp}}$}\\
\hline
\multicolumn{2}{@{}l}{Probabilities observable in original data}\\
\quad$\operatorname{Pr}(Y_2=0)$ in the 1st refreshment sample &0.84\\
\quad$\operatorname{Pr}(Y_3=0)$ in the 2nd refreshment sample &0.40\\
\quad$\operatorname{Pr}(Y_2=0|W_1=1)$ &0.90\\
\quad$\operatorname{Pr}(Y_3=0|W_1=1, W_2=1)$ &0.98\\
\quad$\operatorname{Pr}(Y_3=0|W_1=0, W_2=1)$ &0.93\\
\quad$\operatorname{Pr}(Y_2=0, Y_3=0|W_1=1, W_2=1)$ &0.98\\
\quad$\operatorname{Pr}(Y_2=0, Y_3=1|W_1=1, W_2=1)$ &0.87\\
\quad$\operatorname{Pr}(Y_2=1, Y_3=0|W_1=1, W_2=1)$ &0.92\\[6pt]
\multicolumn{2}{@{}l}{Coefficients in regression of $Y_3$ on}\\
\quad INT&0.61\\
\quad AGE1&0.72\\
\quad AGE2 &0.74\\
\quad AGE3 &0.52\\
\quad COLLEGE&0.89\\
\quad MALE &0.76\\
\quad BLACK &0.90\\
\quad$Y_1$ &0.89\\
\quad$Y_2$&0.84\\
\quad$Y_1Y_2$&0.89\\
\hline
\end{tabular*}
\end{table}

\section{Concluding Remarks}\label{sec6}


The APYN analyses, as well as the simulations, illustrate the
benefits of refreshment samples for diagnosing and adjusting for panel
attrition bias. At the same time, it is important to recognize that
the approach alone does not address other sources of nonresponse 
bias.\vadjust{\goodbreak}
In particular, we treated nonresponse in wave 1 and the refreshment
samples as ignorable. Although this simplifying assumption is the
usual practice in the attrition correction literature (e.g.,
\cite{hirano1998combining};
\cite{bhattacharya2008inference}), it is worth questioning
whether it is defensible in particular settings.
For example, suppose in the APYN survey that
people disinterested in the campaign chose not to respond to the
refreshment samples, for example, because people disinterested in the
campaign were more likely to agree to take part in a political survey
one year out than one month out from the election. In such a scenario,
the models would impute too many interested participants among the
panel attriters, leading to bias. Similar issues can arise with item
nonresponse not due to attrition.


We are not aware of any published work in which nonignorable
nonresponse in the
initial panel or refreshment samples is accounted for in inference.
One potential path forward is to break the nonresponse adjustments into
multiple stages. For example, in stage one the analyst imputes plausible
values for the nonrespondents in the initial wave and refreshment
sample(s) using
selection or pattern mixture models developed for cross-sectional
data (see \cite{littlerubin}). These form a completed data set except
for attrition and missingness by design, so that we are back in the
setting that motivated Sections~\ref{sec3} and~\ref{sec4}. In stage two, the analyst
estimates the
appropriate AN model with the completed data to perform multiple
imputations for attrition (or to use model-based or survey-weighted
inference). The analyst can
investigate the sensitivity of inferences to multiple assumptions
about the nonignorable missingness mechanisms in the initial wave and
refreshment samples. This approach is related to two-stage multiple
imputation (\cite{shen2000}; \cite{rubin2003}; \cite{junedofer})

More generally, refreshment samples need to be representative of the
population of interest to be informative. In many contexts, this
requires closed populations or, at least, populations with
characteristics that do not change over time in unobservable ways. For
example, the persistence effect in the APYN multiple imputation
analysis (i.e., people interested in earlier waves
remain interested in later waves)
would be attenuated if people who are disinterested in the initial wave
and would be so
again in a later wave are disproportionately removed from the
population after the first wave.
Major population composition
changes are rare in most short-term national surveys like the APYN,
although this could be more consequential in panel surveys
with a long time horizon or of specialized populations.


We presented model-based and multiple imputation approaches to
utilizing the information in refreshment samples. One also could
use approaches based on inverse probability weighting. We are not
aware of any published research that thoroughly evaluates the merits of
the various approaches in refreshment sample contexts. The only
comparison that we identified was in \citet{nevo2003using}---which
weights the
complete cases of the panel so that the moments of the weighted data
equal the
moments in the refreshment sample---who briefly mentions towards the end
of his article that the results from the weighting approach and the
multiple imputation in \citet{hirano1998combining} are similar.
We note that \citet{nevo2003using} too has to make identification
assumptions about interaction
effects in the selection model.

It is important to emphasize that the combined data do not provide any
information about the interaction effects that we identify as
necessary to exclude from the models. There is no way around
making assumptions about these effects. As we demonstrated, when the
assumptions are wrong, the additive
nonignorable models could generate inaccurate results. This limitation
plagues model-based, multiple imputation and re-weighting methods.
The advantage of including refreshment samples in data collection is
that they allow
one to make fewer assumptions about the missing data mechanism than if
only the original panel were available.
It is relatively straightforward to perform sensitivity
analyses to this separability assumption in two-wave settings with
modest numbers of outcome
variables; however, these sensitivity analyses are likely to
be cumbersome when many coefficients are set to zero in the
constraints, as is the case with multiple outcome variables or waves.

In sum, refreshment samples offer valuable information that can be used
to adjust
inferences for nonignorable attrition or to create multiple
imputations for secondary analysis. We believe that many longitudinal
data sets could benefit from the use of such samples, although
further practical development is needed, including methodology for
handling nonignorable unit and
item nonresponse in the initial panel and refreshment samples,
flexible modeling strategies for high-dimensional panel data,
efficient methodologies for inverse probability weighting and thorough
comparisons of them to model-based and multiple imputation approaches,
and methods for extending to more complex designs like multiple\break waves
between refreshment samples. We hope that
this article encourages researchers to work on these issues and data
collectors to consider supplementing
their longitudinal panels with refreshment samples.\looseness=1

\section*{Acknowledgments}

Research supported in part by NSF Grant SES-10-61241.



\end{document}